\documentstyle[12pt]{article}

\begin{document}

\begin{center}

{\Large  {\bf Comment on  ``Gauge transformations are Canonical
transformations"}}

\vspace{.5in}

Pathikrit Bhattacharya \footnote{pathikritb@gmail.com} and Bhabani
Prasad Mandal\footnote{bhabani@bhu.ac.in ; bpm@bose.res.in}

Department of Physics, Banaras Hindu University, Varanasi-221005,
INDIA.

\vspace{.9in}

{\bf Abstract}
\end{center}

We comment on the work of Tai L Chow, Eur. J. Phys. 18, 467
(1997). By considering the Lagrangians which  are uniquely defined
only to within an additive total time derivative of a function of
co-ordinates and time the author has tried to show that the gauge
transformations which relate these Lagrangians are canonical
transformations. He has obtained the right conclusion only by
using wrong canonical equations and the entire exercise has hence
become erroneous and inconclusive. By using the definition of
canonical transformation through Poisson brackets we prove that
the above gauge transformations are canonical transformations.

\vspace{.5in}
%\section{Introduction}

We refer here to the letter by Chow \cite{ch} in which the author
has tried to show that the gauge transformations relating  the
Lagrangians differ by a total time derivative of arbitrary
function of coordinate and time are in fact canonical. The
approach however is misleading. The author has obtained the right
conclusion only by using wrong canonical equations [ see the last
two equations of \cite{ch}] and the entire exercise has hence
become erroneous and inconclusive.  In this letter we give a
complete proof that the gauge transformations are canonical
transformations by using the definition of canonical
transformations in terms of Poisson brackets.

The transformation $(p_i,\ q_i)\longrightarrow (P_i, \ Q_i)$ is
canonical iff the following Poisson Brackets are satisfied
\begin{equation}
[P_i,\  P_k]_{q,p} =0 =[Q_i,\ Q_k]_{q,p};\ \& \ [Q_i,\ P_k]_{q,p}
= \delta_{ik}.\label{pb}
\end{equation}
To calculate these Poisson Brackets we consider the Hamiltonians
corresponding to $L(q_i,\dot{q}_i,t)$ and $L^\prime(q_i,\dot{q}_i,
t) $ defined through Legendre transformations
\begin{eqnarray}
H(p_i, q_i,t)&= &\sum_j p_j\dot{q}_j -L(q_i,\dot{q}_i,t)
\nonumber\\ H^\prime(P_i, Q_i,t)&= &\sum_j P_j\dot{q}_j
-L^\prime(q_i,\dot{q}_i,t)
\end{eqnarray}
where $$ P_i = \frac{\partial L^\prime}{\partial \dot{q}_i}= p_i
+\frac{\partial}{\partial\dot{q}_i}\frac{df}{dt}(q_i,t)= p_i
+\frac{\partial}{\partial q_i}f(q_i,t)$$ and $Q_i= q_i $ as
$$\frac{df}{dt}(q_i,t) = \sum_i\frac{\partial f}{\partial
q_i}\dot{q}_i +\frac{\partial f}{\partial t}$$

Now we to show, the transformation $(p_i,q_i)\longrightarrow
(P_i,Q_i)$ is canonical all we need to show that the transformed
canonical variables, $(P_i, Q_i)$ satisfy the Poisson brackets
listed in Eq. (\ref{pb}).
\begin{equation}
[Q_i,\ Q_k]_{q,p} = \sum_j[ \frac{\partial Q_i}{\partial
q_j}\frac{\partial Q_k}{\partial p_j}-\frac{\partial Q_i}{\partial
p_j}\frac{\partial Q_k}{\partial q_j}] =0
\end{equation}
as $\frac{\partial Q_i}{\partial p_j}=0 $ and
\begin{eqnarray}
[P_i,\ P_k]_{q,p} &= &\sum_j[ \frac{\partial P_i}{\partial
q_j}\frac{\partial P_k}{\partial p_j}-\frac{\partial P_i}{\partial
p_j}\frac{\partial P_k}{\partial q_j}]\nonumber \\ &=&
\frac{d}{dt}[  \frac{\partial^2 f}{\partial q_k\partial q_i}-
\frac{\partial^2 f}{\partial q_i\partial q_k}] =0
\end{eqnarray}
and finally
\begin{equation}
[Q_i,\ P_k]_{q,p} = \sum_j[ \frac{\partial Q_i}{\partial
q_j}\frac{\partial P_k}{\partial p_j}-\frac{\partial Q_i}{\partial
p_j}\frac{\partial P_k}{\partial q_j}] = \delta_{ik}
\end{equation}

In conclusion, we have shown that the gauge transformations which
connect the Lagrangians differ by a total time derivative of
co-ordinate and time are canonical.  We have pointed out that
approach of ref. \cite{ch} is erroneous and inconclusive.

\end{document}